\documentclass{article}

\usepackage{PRIMEarxiv}
\usepackage{xcolor}
\usepackage[utf8]{inputenc} 
\usepackage[T1]{fontenc}    
\usepackage[colorlinks]{hyperref}       
\usepackage{url}            
\usepackage{booktabs}       
\usepackage{amsfonts}       
\usepackage{nicefrac}       
\usepackage{microtype}      
\usepackage{lipsum}
\usepackage{fancyhdr}       
\usepackage{graphicx}       
\usepackage{amsmath}
\usepackage[compress]{cite}
\graphicspath{{media/}}     
\hypersetup{
citecolor = blue
}
\title{Laser-Induced Fluorescence Spectroscopy (LIFS) of Trapped Molecular Ions in Gas-phase
\thanks{\textit{\underline{Citation}}: 
\textbf{Hemanth Dinesan and S Sunilkumar. Laser-Induced Fluorescence Spectroscopy (LIFS) of Trapped Molecular Ions in Gas-phase. Pages.... DOI:000000/11111.}} 
}

\author{
  Hemanth Dinesan and S. Sunil Kumar  \\
  Department of Physics and Center for Atomic, Molecular, and Optical Sciences and
Technologies (CAMOST) \\
  Indian Institute of Science Education and Research (IISER) \\
 Tirupati-517507\\
  \texttt{hemanthdinessan@iisertirupati.ac.in} \\
}

\begin{document}
\maketitle

\begin{abstract}
This review presents the Laser-Induced Fluorescence Spectroscopy (LIFS) of trapped gas-phase molecular ions. A brief description of the theory and experimental approaches involved in fluorescence spectroscopy, together with state-of-the-art LIFS experiments employing ion traps, is presented. Quadrupole ion traps are commonly used for spatial confinement of ions. One of the main challenges involved in such experiments is poor Signal-to-Noise Ratio (SNR) arising due to weak gas-phase fluorescence emission, high background noise, and small solid angle for the fluorescence collection optics. The experimental approaches based on the integrated high-finesse optical cavities provide a better (typically an order of magnitude more) SNR in the detected fluorescence than the single-pass detection schemes. Another key to improving the SNR is to exploit the maximum solid angle of light collection by choosing high numerical aperture (NA) collection optics. The latter part of the review summarises the current state-of-the-art intrinsic fluorescence measurement techniques employed for gas-phase studies. Also, the scope of these recent advances in LIFS instrumentation for detailed spectral characterisation of a fluorophore of weak gas-phase fluorescence emission is discussed, considering fluorescein as one example.
\end{abstract}

\keywords{Fluorescence, Quantum yield, Lifetime, Quadrupole ion traps, Fluorescein}

\section{Introduction}
One of the most significant innovations in biology is the discovery of fluorescent probes, which enabled the scientific community to interpret various biological phenomena that remained un-investigated till then. It all started with the accidental discovery of mauveine (1856), the first organic chemical, and the synthesis of fluorescein (1871), one of the most widely employed fluorophores today \cite{cova2017reconstructing, warrier2014fluorescent}. Since then, numerous fluorescent probes have been synthesised and used extensively. Currently, fluorescent probes are routinely used for research due to their wide range of biomedical applications, such as pharmaceutical research, clinical research, diagnostics, etc. \cite{warrier2014fluorescent}. Making use of their unique photoluminescence properties, these probes enable fast response and highly sensitive and non-destructive on-site analysis of specific targets. They also outperform the traditional techniques like titrimetry, chromatography, electrochemistry, chemiluminescence, and flow injection analysis \cite{gonccalves2009fluorescent, han2010fluorescent, patsenker2008fluorescent}. Out of many currently available fluorophores, the xanthene-based dyes rhodamine, fluorescein, and their derivatives, are preferred due to their potential “in vivo” applications \cite{warrier2014fluorescent, jiao2018novel, robertson2013fluorescein, o2006safety}. Despite all these applications, a detailed understanding of the intrinsic fluorescence properties of some of these dyes, fluorescein, in particular, is still lacking \cite{yao2013fluorescence, yao2011infrared, tanabe2012molecular, kjaer2020gas}. A complete understanding of the process by which photon energy is redistributed within the molecule in the gas phase and solvent phase requires both experimental and complementary theoretical investigations of the system in the gas phase as well as in a controlled solvent environment. The knowledge one may acquire using this model system could enable one to manipulate the fluorescence properties of similar molecules and design molecules of predictable fluorescence properties.

Fluorescence spectroscopic experiments are routinely performed to monitor molecules/ions in the condensed phase, while the gas-phase measurements require specialised arrangements. Recent developments in ion-trap-based spectroscopic measurements have made it possible to isolate and individually probe the behaviour of the fluorophore of interest \cite{yao2013fluorescence, bian2010gas, svendsen2017origin}. Most of these experimental approaches involve a combination of laser-induced fluorescence measurement and ion-trap mass spectrometry. Briefly, the intrinsic fluorescence studies involve isolating the ions from their natural environment by ion-trap mass spectrometric techniques, probing them with laser-induced fluorescence spectrometers, and analysing the fluorescence emission spectra to determine the fluorescence lifetimes and fluorescence efficiencies. Often the fluorescence measurement in the gas phase is quite challenging due to poor Signal-to-Noise Ratio (SNR) limited by several factors, namely, (i) weak fluorescence emission of some molecular species in the gas phase, (ii) the limited number of ions that can be stored within an ion trap, (iii) background noise arising from the scattering of the exciting beam, and (iv) small solid angle for fluorescence collection optics due to the poor accessibility of the ions in closed trap geometries.
This review article presents an overview of the state-of-the-art techniques in laser-based fluorescence experiments for the intrinsic fluorescence measurements of trapped molecular ions. We address various measurement techniques adopted so far, their pros and cons, and present an application of these techniques by reporting the state-of-the-art measurements on fluorescein.

This review is organised as follows. We start with a brief description of the theory and generally adopted experimental approaches in fluorescence spectroscopy. The following section provides an overview of the state-of-the-art experiments performed for measuring the fluorescence from gas-phase molecular species. Most of these studies involve ion traps. This section discusses various approaches based on single-pass and cavity-integrated schemes. The review then focuses on applying the LIFS techniques in the intrinsic fluorescence studies of xanthene-based dyes, with a particular emphasis on fluorescein, followed by a discussion on the scope for possible future improvements. The last section provides a summary and conclusion.

\section{Background: Theory and experimental approaches}
\begin{figure}
  \centering
  \includegraphics[width=.6\linewidth]{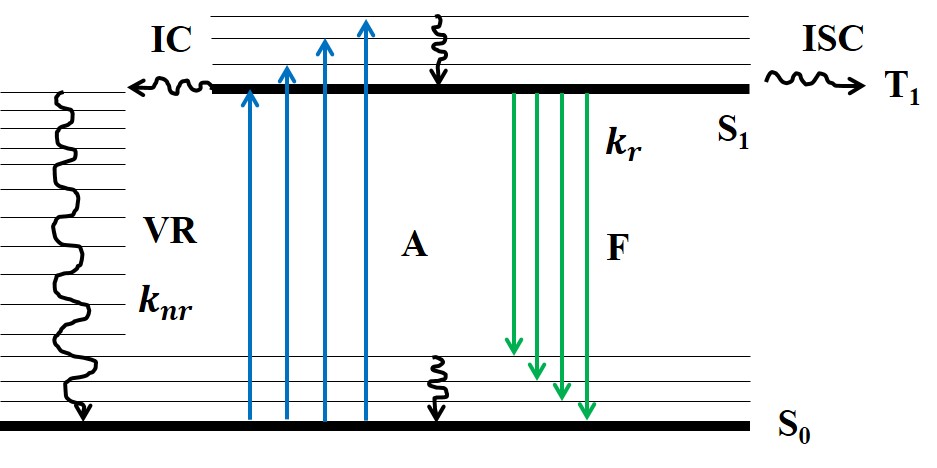}
  \caption{A simplified form of the Perrin-Jablonski diagram illustrating some of the important processes involved in the interaction of a molecule with light. $S_0$ and $S_1$ are the ground and excited electronic singlet states, respectively. Absorption and fluorescence are radiative processes denoted by A (blue) and F (green), while the processes like vibrational relaxation (VR), intersystem crossing (ISC), internal conversion (IC), etc., are also illustrated. $k_r$ is the radiative decay rate and $k_{nr}$ is the non-radiative decay rate. Phosphorescence is out of the scope of this paper (not shown), which involves the triplet state ($T_1$).}
  \label{figpjd}
\end{figure}

The mechanism of fluorescence is illustrated in figure \ref{figpjd} with the help of a simplified Perrin-Jablonski diagram indicating light-matter interaction and molecular energy levels involved in the fluorescence process \cite{valeur2012molecular}. The energy level diagram shown corresponds to a molecular species having no unpaired electrons forming a singlet ground state (which is also applicable for charged systems formed by protonation or deprotonation of a molecule having an even number of electrons). In a fluorescence experiment, light from a source such as a lamp or a laser illuminates a molecule that absorbs the photon energy and gets excited from the ground electronic singlet state $S_0$ to an excited electronic singlet state $S_1$.
Once the molecule reaches the excited state, it subsequently returns to the ground state via various processes, either radiative or non-radiative. Fluorescence emission is a radiative process in which the emitted photon typically has a longer wavelength than the absorbed photon. Two commonly measured parameters in fluorescence experiments are fluorescence quantum yield and fluorescence lifetime. In the two subsequent sections, we provide a brief description of how these quantities are conventionally defined and measured, which mainly pertain to solvent-phase experiments. These descriptions are provided to highlight their difference from gas-phase techniques that we focus on in later sections.
\subsection{Fluorescence quantum yield ($\Phi_f$)}
Fluorescence quantum yield is defined as the ratio of the number of photons emitted to the number of photons absorbed by the molecule or ion after being shined by an excitation light source. Quantum yield is given by,
\begin{equation} \label{eqphif}
\Phi_f = \frac{k_r}{k_r+ k_{nr}}
\end{equation}
where $k_r$ is the radiative decay rate and $k_{nr}$ is the non-radiative decay rate \cite{valeur2012molecular}. If the non-radiative decay rate is negligible compared to the radiative decay rate, then $\Phi_f$ is close to unity.  Fluorescence quantum yields are often affected by the characteristics of the environment of the molecule, such as polarity, pH, and interactions with amino acids or nucleobases. Hence the precise determination of intrinsic quantum yields and their dependence on the environment is critical for several applications. Fluorescence quantum yield measurements can be performed either by absolute measurements or relative to the reference material of known quantum yield. Absolute measurement is the only way to determine the quantum yields of transparent samples absorbing or emitting in the near-infrared region where no reliable standard materials of known quantum yields are available. Absolute fluorescence quantum yield measurements are usually performed by the integrated sphere approach and are given by the following expression \cite{kawamura2004simple}
\begin{equation}\label{eqphif2}
\Phi_f = \frac{N_{Em}}{N_{Abs}} = \frac{{\int{I_{Em}(\lambda)}\,d\lambda}}{\int{{[I_{Ex}{(\lambda)}- I^\prime_{Ex}(\lambda)}]\,d\lambda}}
\end{equation}
where $ N_{Em}$ is the number of photons emitted by the sample, $N_{Abs}$ is the number of photons absorbed by the sample, $\lambda$ is the wavelength, $I_{Em}(\lambda)$ is the fluorescence intensity of the sample, $I_{Ex}(\lambda)$ is the intensity of the excitation source without the sample and $I^\prime_{Ex}(\lambda)$ is the intensity of the excitation laser in the presence of the sample.

Relative fluorescence quantum yield measurements are usually performed by Spectrofluorometer and is given by the following equation \cite{levitus2020tutorial}
\begin{equation} \label{eqphisr}
\frac{\Phi_f^S}{\Phi_f^R} = \frac{n_S^2}{n_R^2} \times F_I \times \frac{\left(1-10^{-A^R(\lambda_{Ex})}\right)}{\left(1-10^{-A^S(\lambda_{Ex})}\right)}
\end{equation}
\begin{center}
\begin{equation*}
F_I = \frac{\int_{0}^{\infty}{I^S_f (\lambda_{Ex}, \lambda_{Em}) \,d\lambda_{Em}}}{\int_{0}^{\infty}{I^R_f (\lambda_{Ex}, \lambda_{Em}) \,d\lambda_{Em}}}
\end{equation*}
\end{center}
In equation \ref{eqphisr}, the superscripts $S$ and $R$ refer to sample and reference, respectively. $\Phi_f$ is the fluorescence quantum yield, $n$, the refractive index, $I_f$, the fluorescence intensity, and A, the absorbance expressed as $A=-\log(I_T/I_0)$, with $I_0$ and $I_T$ representing the incident and transmitted intensities, respectively. $F_I$ is the fluorescence intensity given as the ratio of the total number of detected photons at all wavelengths for the sample and the reference. It is obtained from the corrected emission spectra. The term $1-10^{-A(\lambda_{Ex})}$ is the fraction of the incident photons that are absorbed by the sample/reference, as indicated by the superscript on $A$, at a given excitation wavelength which can be approximated as the area under the absorption curve for
low values of absorbance ($A\approx 0.025$). The value of the fluorescence quantum yields can be retrieved by measuring the areas under the absorption and emission curves, knowing the refractive indices values of the sample and reference, and taking the value of $\Phi^R_f$ from the IUPAC database \cite{brouwer2011standards, magde2002fluorescence}.
\subsection{Fluorescence lifetime ($\tau$)}
Fluorescence lifetime ($\tau$) is defined as the average time a molecule spends in the excited state ($S_1$) prior to its return to the ground state ($S_0$) by emitting a fluorescence photon \cite{valeur2012molecular}. Generally, fluorescence lifetimes are less than 10 ns. It is given by,
\begin{equation} \label{eqtau}
\tau = \frac{1}{k_r + k_{nr}}
\end{equation}
where $k_r$ is the radiative decay rate and $k_{nr}$ is the nonradiative decay rate. In the absence of nonradiative processes $k_{nr}$ becomes zero and this lifetime is called the \emph{intrinsic} or \emph{natural fluorescence lifetime} ($\tau _n$) and can be linked to the fluorescence quantum yield as
\begin{equation} \label{eqtaun}
\tau_n = \frac{\tau}{\Phi_f}
\end{equation}
Hence, we can retrieve the natural fluorescence lifetime by measuring the fluorescence lifetime and quantum yield.

The experimental approach widely adopted to determine $\tau$ is the pulse lifetime measurement method. During this process, the sample is excited with a laser pulse, and the fluorescence intensity is measured as a function of time. The time required for the fluorescence intensity to decay to ${1/e}$ times its initial value yields the $\tau$ value. To perform this operation, one has to overcome many challenges. The main difficulty is using excitation lasers with picosecond pulse widths to measure the lifetimes of a few nanoseconds. Further, the response time of the detecting unit (generally photomultiplier tubes) must be in the sub-nanoseconds range. Also, while exciting the sample with repetitive pulses, one should be extremely careful to set the gap between two consecutive pulses to be at least $4-5$ times greater than the fluorescence lifetime. This is required to avoid overlap of the fluorescence signal induced by one pulse to the response caused by the previous pulse. The time-resolved fluorescence can be then measured with photon-counting methods. As the name indicates, it involves counting the number of fluorescence photons reaching the photon-counting device at different instances after the excitation pulse. This gives a decay curve that can be fitted with an exponential decay function or a sum of exponential decay functions for accurate lifetime retrievals. The non-exponential nature of the fluorescence lifetime distribution may arise due to several reasons, namely, (i) fluorescence quenching due to the fluorophore-solvent molecule interaction, (ii) contribution of fluorescence emission from more than one singlet excited state, and (iii) the presence of different prototropic forms of the molecule in the solution. It may also arise due to the competition with other fast relaxation processes when the fluorescence decay rate falls in the picoseconds range. Fluorescence decay may also exhibit a non-exponential character when working with heterogeneous samples, i.e., samples with different fluorophores or the same fluorophores in different environments.
\subsection{Ion Traps}

\begin{figure}[ht]
\begin{center}
\includegraphics[width=.4\linewidth]{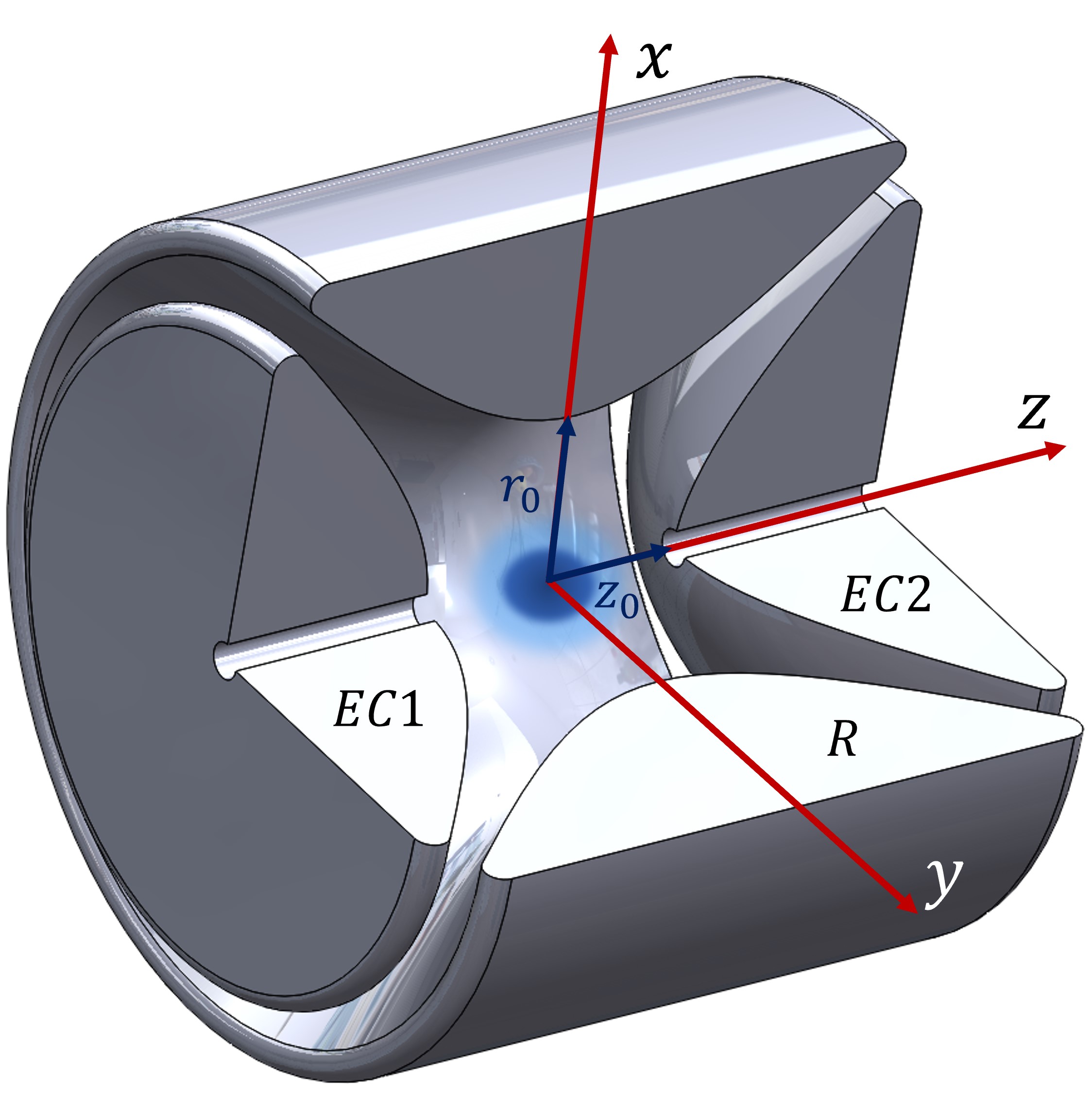}
\caption{A cross-sectional view of a Paul trap. It comprises a hyperbolically-shaped ring electrode and two hyperbolic electrodes having rotational symmetry. EC stands for the endcap electrodes and R for the ring electrode. $r_0$ is the distance from the trap centre to the ring electrode, and $z_0$ is the distance from the trap centre to the EC electrodes. To have an ideal quadrupole potential within the trap, the dimensions are chosen so that it satisfies the relation ${r_0}^2 = 2{z_0}^2$. The holes on the endcap electrodes are used to load and extract the ions.}
\label{figpt}
\end{center}
\vspace{5mm}
\end{figure}

Ion traps are mechanical devices designed to spatially confine atomic and molecular ions for a sufficiently long time so that they can be manipulated to extract desired information. Quadrupole ion traps (QIT) are routinely employed for gas-phase fluorescence measurements. They fall in the category of multipole radiofrequency (RF) ion traps whose working principle is briefed here. QITs are usually built in two main configurations: linear and 3-dimensional. The latter is often referred to as the Paul trap (after its inventor) \cite{paul1990electromagnetic}. It is comprised of a ring electrode and two rotationally symmetric electrodes (refrerred to as the endcaps) as shown in figure \ref{figpt}. Both these electrodes have a hyperbolically-shaped inner surface.
A confinement potential well is formed within the trap when a radiofrequency potential of suitable frequency and amplitude is applied between the ring electrode and the endcap electrodes. A general quadrupole potential is given by, \cite{march2009quadrupole}
\begin{equation} \label{eqVq}
V(x,y,z) =A(\lambda x^{2}+\sigma y^{2} + \gamma z^{2}) + C,
\end{equation}
where $A$ depends on the electric potential applied between the electrodes (an RF potential or a combination of RF and DC potentials) and is independent of $x$, $y$, and $z$. $C$ is the DC potential applied to all electrodes to float the trap and $\lambda$, $\sigma$, and $\gamma$ are constants. On applying the Laplace equation $\nabla^{2}V = 0$, equation \ref{eqVq} becomes
\begin{equation} \label{eqqlin}
2A(\lambda+ \sigma + \gamma)=0
\end{equation}

For a linear QIT, $\lambda = -\sigma = 1$ and $\gamma = 0$, and for a Paul trap, $\lambda = \sigma = 1$ and $\gamma = -2$. For a Paul trap in cylindrical polar coordinates, equation \ref{eqVq} takes the form
\begin{equation} \label{eqq3d}
V(x,y,z)= A(r^2 - 2z^2) +C
\end{equation}

Within a Paul trap, trapped ions are localised in a tiny volume, which favours a high fluorescence signal collection efficiency. Therefore, the Paul trap design has been employed by several groups for performing fluorescence spectroscopy \cite{khoury2002pulsed,bian2010gas, sagoo2011fluorescence, sassin2009fluorescence, tiwari2020breaking, danell2003fret}. Paul traps possess a clear advantage of being able to confine the ion cloud at the trap-centre with a near spheroid geometry and hence the fluorescence emission can be considered to arise from a point source. The Penning trap \cite{brown1986geonium} features a geometry somewhat similar to that of a Paul trap, except that the applied potentials are not time-varying and that a strong magnetic field generated with the help of a pair of superconducting magnets is used to confine the motion of the ions in the $x-y$ plane. Zenobi and co-workers have employed the Penning trap geometry for performing the gas-phase fluorescence spectroscopy \cite{frankevich2005laser,chingin2009exploring}. Since a Penning trap does not require an RF field, it does not suffer from the issue of RF heating \cite{gerlich1992inhomogeneous}. However, Paul and Penning traps suffer from limited optical access and moderate light collection efficiency due to the ring electrode geometry ($\sim 2\%$, even with the modifications introduced in the state-of-the-art instruments as discussed later)  \cite {tiwari2020breaking, bian2010gas}. 

\begin{figure}[ht]
\begin{center}
\includegraphics[width=.6\linewidth]{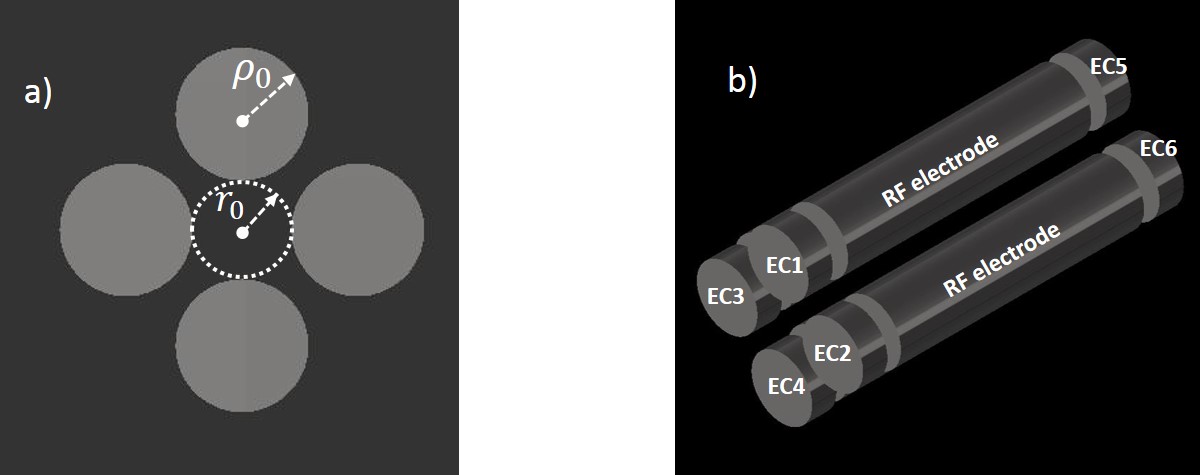}
\caption{a) Cross-section of a quadrupole ion trap. It consists of four metal rods of radius $\rho_0$ arranged in a cylindrical symmetry with an inscribed radius $ r_0$. b) 3D view of the same in which the endcaps on either side of quadrupole rods which act as entrance and exit electrodes, are shown. These endcaps (denoted as $EC_x$, where $x$ = 1,2...8) are supplied with DC potentials and are opened and closed for ion loading into the trap and ion extraction from the trap while a radiofrequency (RF) sinusoidal voltage is applied to the RF electrodes.}
\label{figql}
\vspace{5mm}
\end{center}
\end{figure}

The limitation in the optical access encountered by the Paul trap and Penning trap geometries can be minimised by employing a linear quadrupole ion trap, a schematic representation of which is shown in figure \ref{figql}. A custom geometry of a linear quadrupole ion trap with an expanded inscribed diameter for providing improved optical access has been demonstrated by Vaishnavi Rajagopal et al. for performing gas-phase fluorescence spectroscopy \cite{rajagopal2017linear}. Even if this design provides a larger trapping volume and improved optical access compared to the Paul and Penning traps, it is not very well-suited for a high fluorescence collection efficiency. First, the size of the long cylindrical trapped ion cloud geometry makes it difficult to access all the stored ions with a laser beam. Secondly, it is mandatory to have a large NA lens for fluorescence collection.
Such a trap is formed of 4 rods of radius $\rho_0$ arranged in a cylindrical symmetry of inscribed radius $r_0$ where the diameter of the rods is chosen to satisfy the relation $\rho_0= 1.148 r_0$ \cite{wester2009radiofrequency, gerlich1992inhomogeneous}. This choice is made to generate a potential closest to an ideal quadrupole potential within the trap. The rods are supplied with a radiofrequency (RF) voltage such that the neighbouring electrodes have opposite polarity. The time-varying electric potential created within a multipole radiofrequency ion trap of pole order $n$ ($n=2$ for a quadrupole) is given in cylindrical coordinates by
\begin{equation} \label{eqVr}
V (r,\phi,t) = V_0\cos(n\phi)\times\left(\frac{r}{r_0}\right)^n\times\sin(\omega t)
\end{equation}
where $V_0$ is the radiofrequency amplitude and $\omega$ denotes the angular frequency. The effective potential that will radially confine the ions within the trap is given by \cite{gerlich1992inhomogeneous},
\begin{equation} \label{eqVeff}
V_{eff}(r) = \frac{ n^2q^2V_0^2}{4m\omega^2r_0^2}\times\left(\frac{r}{r_0}\right)^{2n-2}
\end{equation}
where $r$ is the co-ordinate of drift motion of ions of mass $m$ and charge $q$ in the trap. Usually, in QIT-based experiments, the ions generated by electrospray ionisation source or by matrix-assisted laser desorption/ionisation (MALDI) source with kinetic energies of a few electronvolts are guided with ion optics, mass-selected using an RF quadrupole mass filter and sent into the ion trap. Ions are loaded into the trap by lowering the potentials of the entrance endcaps (denoted as EC1, EC2, EC3, and EC4 in figure \ref{figql}(b)) and raising it after a sufficient number of ions are loaded. The trapped ions are cooled with an inert buffer gas maintained at a well-controlled temperature \cite{choi2012effective}. After a selected storage duration within which the ions may be subjected to laser irradiation for fluorescence measurements, the exit endcap potentials (denoted as EC5, EC6, EC7, and EC8) are lowered, and the ions are extracted for characterisation, for instance, by time-of-flight mass spectrometry \cite{wiza1979microchannel}. 

The laser-induced fluorescence spectroscopy of a trapped ion cloud can be performed to study the electronic structure of atomic and molecular species \cite{blatt1982precision, danon1982laser, wang2001direct}. In essence, the trapped ion cloud is probed by a laser beam of the desired wavelength, and the fluorescence emission resulting from the photon-ion interaction is collected by dedicated optics and a photodetector. A brief overview of the conventional experimental approaches adopted in fluorescence experiments is outlined in the following section.
\subsection{Approaches for fluorescence detection}
\begin{figure}[ht]
\begin{center}
\includegraphics[width=.4
\linewidth]{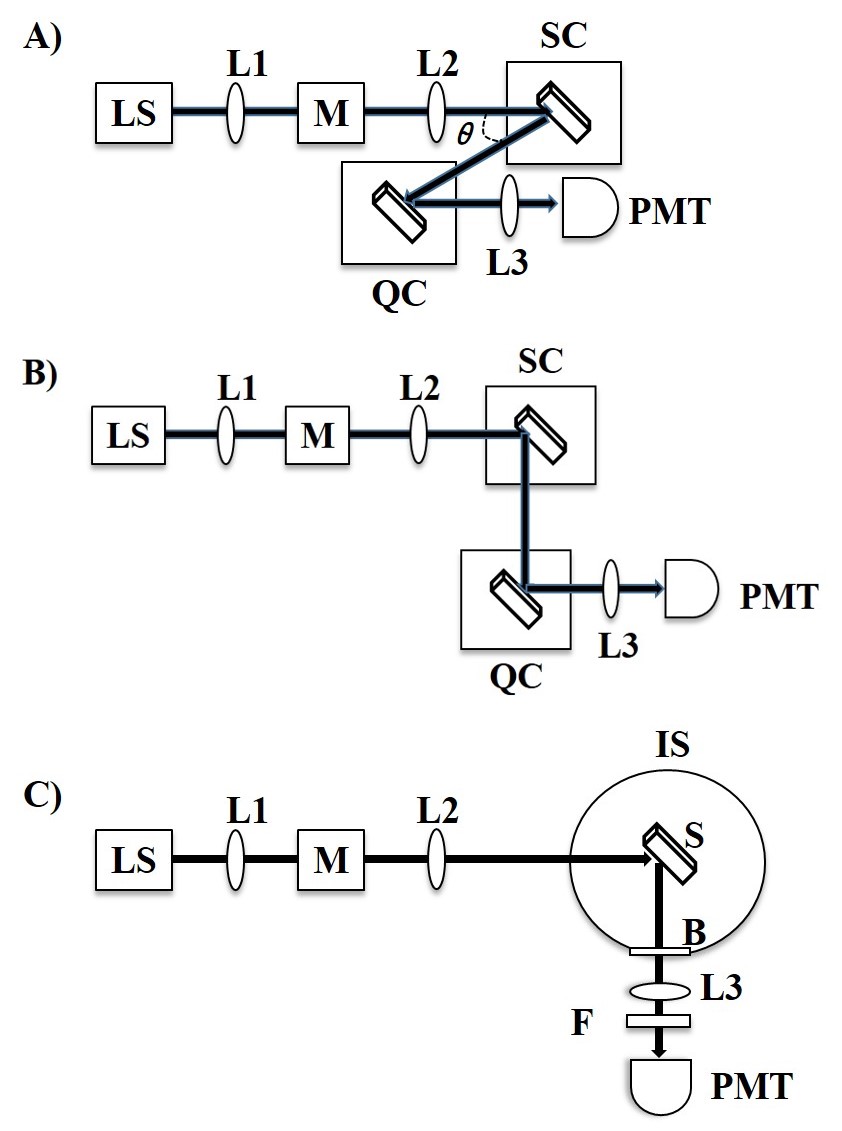}
\caption{Schematic of the optical setups used for fluorescence experiments. a) Vavilov-Melhiush approach involving the front-face excitation and detection geometry, b) Weber-Teale approach involving front-face sample excitation and detection perpendicular to the excitation beam direction, and c) Integrating sphere approach. LS stands for excitation source, L; lens, M; monochromator, S; sample, SC; sample cuvette, QC; quantum counter, IS; integrating sphere, B; baffle, F; filter and PMT is the photomultiplier tube used for fluorescence detection.}
\label{figlif}
\vspace{5mm}
\end{center}
\end{figure}

The most simplified approach for the LIFS consists of an excitation source, typically a lamp or a laser, depending on the requirements, the light from which is sent to the probe molecule located in a cuvette or a cabin or an ion trap.
The resulting emission is collected by a photodetector, typically a photomultiplier tube (PMT) or a gated charge-coupled device (CCD) camera either at a right angle to the excitation beam (Weber-Teale approach) \cite{weber1957determination, demasa1968measurement} or at acute angles to the excitation beam (Vavilov-Melhiush approach) \cite{vavilov1924fluorescence, melhuish1972absolute}. The basic layout of commonly adopted approaches is depicted in figure \ref{figlif}.
Figure \ref{figlif}(a) depicts the Vavilov-Melhiush approach, which involves front-face excitation of the sample placed in a cuvette and performing fluorescence detection at acute angles to the excitation beam. Later on, the Weber-Teale approach was introduced, which is the conventional geometry adopted for many fluorescence experiments and is shown in figure \ref{figlif}(b) \cite{weber1957determination, demasa1968measurement}. This involves front-face excitation of the sample and performing fluorescence detection at right angles to the excitation direction. The main advantage of this geometry compared to the Vavilov-Melhiush method is that it reduces considerably the stray light originating from the excitation source. This is critical in experiments where the excitation wavelength is within the emission band. The integrating sphere approach consists of irradiating the sample placed in a sphere with an optical fibre-coupled laser source, as shown in figure \ref{figlif}(c). The inner face of the sphere is coated with a diffusely reflecting material ($\mathrm{{BaSO_4}}$), and the photons entering the sphere are redirected to the exit port after being reflected at the inner surface of the sphere \cite{rohwer2005measuring, porres2006absolute}. This light may contain scattered components at the excitation wavelength, which are filtered by long-pass filters, and the fluorescence is detected by dedicated detectors. Out of the three approaches mentioned here, the Weber-Teale approach is employed for the fluorescence measurements from the ion-traps. \cite{bian2010gas,tiwari2020breaking,stockett2016cylindrical}. The Vavilov-Melhiush and integrating sphere approaches are not preferred since the former technique may contain backscattered excitation wavelength components in the detected light while the latter is practically difficult to implement due to constraints in ion-trap geometries.

\subsection{Time-correlated Single-photon Counting (TCSPC)}
TCSPC \cite{o32time} is a technique based on detecting single photons of the fluorescence emission. This method is preferred for experiments where the light levels to be detected are very low, for instance, the gas-phase fluorescence measurements of trapped molecular ions. In such cases, the probability of detecting one photon over one signal period is far less than one \cite{becker2012fluorescence,becker2005advanced}. Time-domain measurements involve exciting the sample with repetitive laser pulses of duration shorter than its fluorescence lifetime and measuring the arrival times of photons after the excitation pulse. 
This procedure is repeated by scanning the laser, focused at various positions on the sample. By recording the arrival times of photons for each position, a histogram is obtained to form a 3D array consisting of the arrival time distribution of photons as a function of the spatial coordinates ($x,y$). The fluorescence decay parameters are obtained by fitting the time distributions with a model, which is a convolution between a suitable decay function and the instrument response function (IRF) \cite{becker2012fluorescence}. This technique possesses many advantages like high time resolution, best lifetime accuracy, and tolerance to the dynamic changes in the fluorescence parameters during the data acquisition. It has numerous applications in fluorescence resonance energy transfer (FRET), scanning mammography, brain imaging, and examining the influence of solvent polarities on fluorophore lifetime, among many others \cite{wurth2013relative, becker2005advanced}. 

\section{State-of-the-art LIFS experiments towards intrinsic fluorescence studies}

A profound understanding of the photochemistry of fluorophores requires a thorough investigation of their spectral properties in the solvent and gas phases by both experimental approaches and complementary theoretical calculations. An extensive amount of work (both theoretical and experimental) was carried out to investigate the photochemistry of several fluorescing molecules/ions over the last two decades \cite{yao2013fluorescence, bian2010gas, friedrich2004time, forbes2011gas, magde2002fluorescence, sagoo2011fluorescence, sassin2009fluorescence, lefebvre2004spectra} thanks to the developments in the mass spectrometry and the advent of pulsed lasers along with the second harmonic generation techniques. Most of the time-dependent fluorescence experiments rely on tunable solid-state lasers operating in the pulsed mode, capable of generating picosecond or femtosecond pulses of high energies. These lasers operating in the near-infrared region are then frequency-doubled by second harmonic generation (SHG) schemes involving non-linear optic crystals \cite{bian2010gas, yao2013fluorescence, nagy2012fluorescence, khoury2002pulsed, tiwari2020breaking, kjaer2020gas}. The SHG laser outputs fall in the visible region where the excitation bands of most dyes mentioned in this review overlap. These SHG techniques are an integral part of most of the time-dependent fluorescence experiments since commercial femtosecond lasers are not readily available in the visible region.
\begin{figure}[ht]
\begin{center}
\includegraphics[width=.4\linewidth]{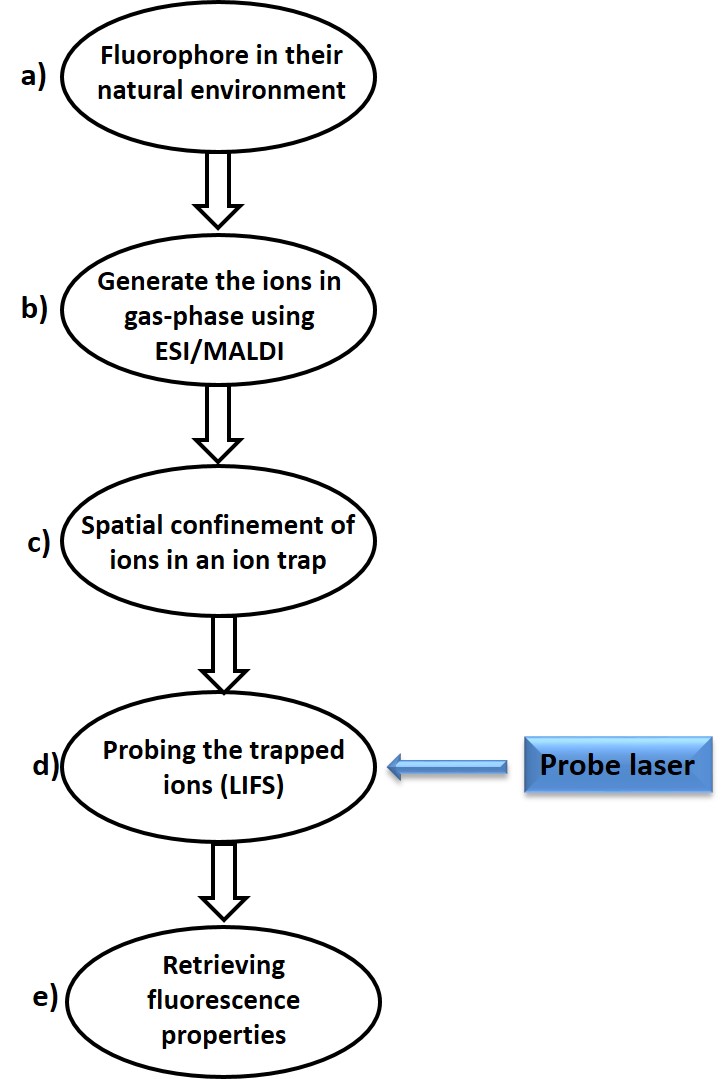}
\caption{Flowchart showing the generally adopted step-by-step process for the intrinsic fluorescence spectroscopic studies. The transition from step a) to c) involves sample preparation, generating gas-phase ions using electrospray ionisation (ESI) or matrix-assisted laser desorption/ionisation (MALDI), and trapping them. d) consists of fluorescence emission measurements upon excitation with a probe laser system, and e) involves the spectral analysis, thereby retrieving parameters like quantum yield and fluorescence lifetime.}
\label{figfss}
\vspace{5mm}
\end{center}
\end{figure}

A generally adopted procedure for performing intrinsic fluorescence studies is illustrated in figure \ref{figfss}. Fluorophores of interest are first isolated from their natural environment using soft ionisation techniques such as electrospray ionisation (ESI) \cite{fenn1990electrospray, dole1968molecular} and matrix-assisted laser-desorption/ionisation (MALDI) \cite{schnolzer1997identification, cole2011electrospray}. These techniques ensure that the fragile molecules are intact during their conversion to gas-phase ions.
In the electrospray ionisation method, the sample is dissolved into a suitable solvent, or a mixture of solvents which is then subjected to a high electric field to generate isolated gas-phase molecular ions \cite{manisali2006electrospray, yamashita1984electrospray}. In MALDI, sample molecules embedded in a matrix are liberated through the energy that is deposited on the matrix using a laser \cite{frankevich2005laser}. These ions are mass-selected and stored in an ion storage device such as a quadrupole ion trap and are then probed by a laser of a suitable wavelength. Ion-trap-based LIFS apparatus, in general, involves a laser source that irradiates the trapped ion cloud along the trap axis over a few tens of seconds (depending on the storage times in the trap). The resulting emission is collected orthogonal to the incident direction by a lens placed very close to the ion cloud to maximise the solid angle of light collection. The collected light is then passed through a long-pass filter to suppress the short wavelength photons from scattered light and a focusing lens to couple the fluorescence photons to the detector, usually a CCD camera or a PMT.

The state-of-the-art approaches adopted to perform fluorescence spectroscopy of several molecular ions in the gas phase are discussed below.
\subsection{Single-pass fluorescence spectroscopy}
In recent years, a significant amount of scientific work has been published which addresses the intrinsic photophysical properties of gas-phase chromophores in ionic form by integrating ion-trap mass spectrometers with laser-based fluorescence detection schemes \cite{yao2013fluorescence, bian2010gas,forbes2011gas, sagoo2011fluorescence, sassin2009fluorescence}. All these works focused on molecular ions of high fluorescence quantum yield, such as xanthene-based dyes, rhodamine, and its derivatives. Rebecca A Jockusch and coworkers demonstrated an apparatus for performing fluorescence measurements from an ion cloud trapped in a quadrupole ion trap (QIT) \cite{bian2010gas}.
\begin{figure}[ht]
\begin{center}
\includegraphics[width=0.6\linewidth]{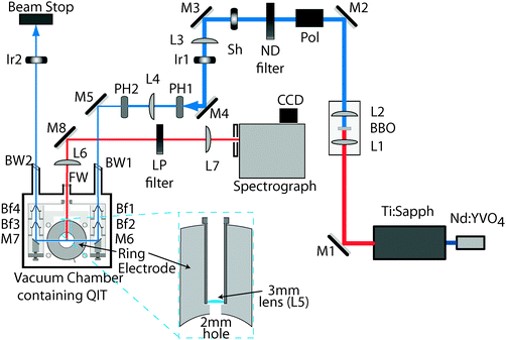}
\caption{Schematic of the setup used by Bian et al. \cite{bian2010gas}. The gas-phase ions trapped in a quadrupole ion trap are illuminated with a tunable laser with an emission wavelength of 493 nm. The resulting fluorescence spectra are collected orthogonally to the incident beam direction with a collimating lens and filter assembly and recorded on a CCD detector. Ti: Sapph stands for Titanium-Sapphire laser, M1 to M8; mirrors, L1 to L7; lenses, BBO; Beta-$\mathrm{BaB_2O_4}$ crystal used for frequency doubling the Ti: Sapph laser output, Pol; polarising beam splitter, ND; neutral density filter, Sh; optical shutter to control the laser irradiation of the QIT, Ir; iris, PH; pinhole, $\mathrm{Nd: YV0_4}$; Neodymium-doped yttrium orthovanadate crystal, BW; Brewster window, Bf; optical baffle, FW; fluorescence window and LP stands for the low-pass filter. \textit{Reprinted from \cite{bian2010gas} Physical Chemistry Chemical Physics \copyright 2010 the Owner Societies}.}\label{figbian}
\vspace{5mm}
\end{center}
\end{figure}
They adopted the conventional procedure as illustrated in figure \ref{figfss}. Figure \ref{figbian} shows a schematic of their experimental setup. After being generated in an electrospray ionisation source, the ions were trapped in a QIT at the centre of the ring electrode. They are then shined by a frequency-doubled Titanium-Sapphire laser whose output wavelength was tuned to the excitation peak of the molecule of interest (493 nm for gas-phase rhodamine, for instance). Two holes drilled on the ring electrode on-axis with the incident beam provided a long optical path for the laser to interact with the trapped ions. The resulting fluorescence was collected through a third hole drilled orthogonal to the incident beam direction. The collected light was collimated, filtered, and spectrally corrected, and the fluorescence spectra were recorded with a CCD camera. Even after subtracting the background noise and the instrument dark noise, the acquired emission spectra may have several scattered components other than fluorescence such as Rayleigh scattering and Raman scattering. Rayleigh scattering component is an elastic scattering component that is present at the excitation wavelength and hence can be filtered by using a cut-off long-pass filter which filters out that part of the spectra while the Raman component is more likely to be present around the fluorescence emission region. The fact that the Raman peaks significantly depend on the excitation wavelength, unlike the fluorescence peaks, makes it possible to distinguish and successfully eliminate their contribution \cite{albrecht2008joseph}. Their geometry provided an overall fluorescence collection efficiency of 0.15\%. They successfully measured the gas-phase excitation and emission spectra of rhodamine 540, 575, 590, and 6G. An optimum choice of laser power of 5 mW and an irradiation time of 20 s was made to minimise other decay processes like photodetachment and fragmentation. The major limitation of their instrument was the low fluorescence collection efficiency, mainly due to a small solid angle of collection (0.25\%) \cite{yao2013fluorescence, bian2010gas, sagoo2011fluorescence, friedrich2004time, forbes2011gas}.

Sassin et al. \cite{sassin2009fluorescence} studied the fluorescence and photodissociation of rhodamine 575 cations in the gaseous phase by a QIT/Time of Flight (QIT/TOF) mass spectrometer setup, which was similar to the instrument mentioned in the previous works \cite{bian2010gas, forbes2011gas} except that a TOF detection scheme was integrated. The trapped ions were irradiated by an argon laser beam at 488 nm for $4-5$ seconds through holes drilled on ring electrodes, while the resulting fluorescence was measured by counting the number of photons emanating from the trapped ion cloud through a third hole drilled on the ring electrode, orthogonal to the incident beam direction. After the irradiation time, the trapped ions were extracted and analysed by a TOF mass spectrometer. The measured decay rates from these methods (photon counting and mass spectrometry) showed good agreement, demonstrating how these two independent techniques can determine the extent of photoactivation and deactivation processes. The main limitation of this setup was also the small solid angle through which the photons were detected due to the geometry of the ring electrode of the trap.

\begin{figure}[ht]
\begin{center}
\includegraphics[width=.6\linewidth]{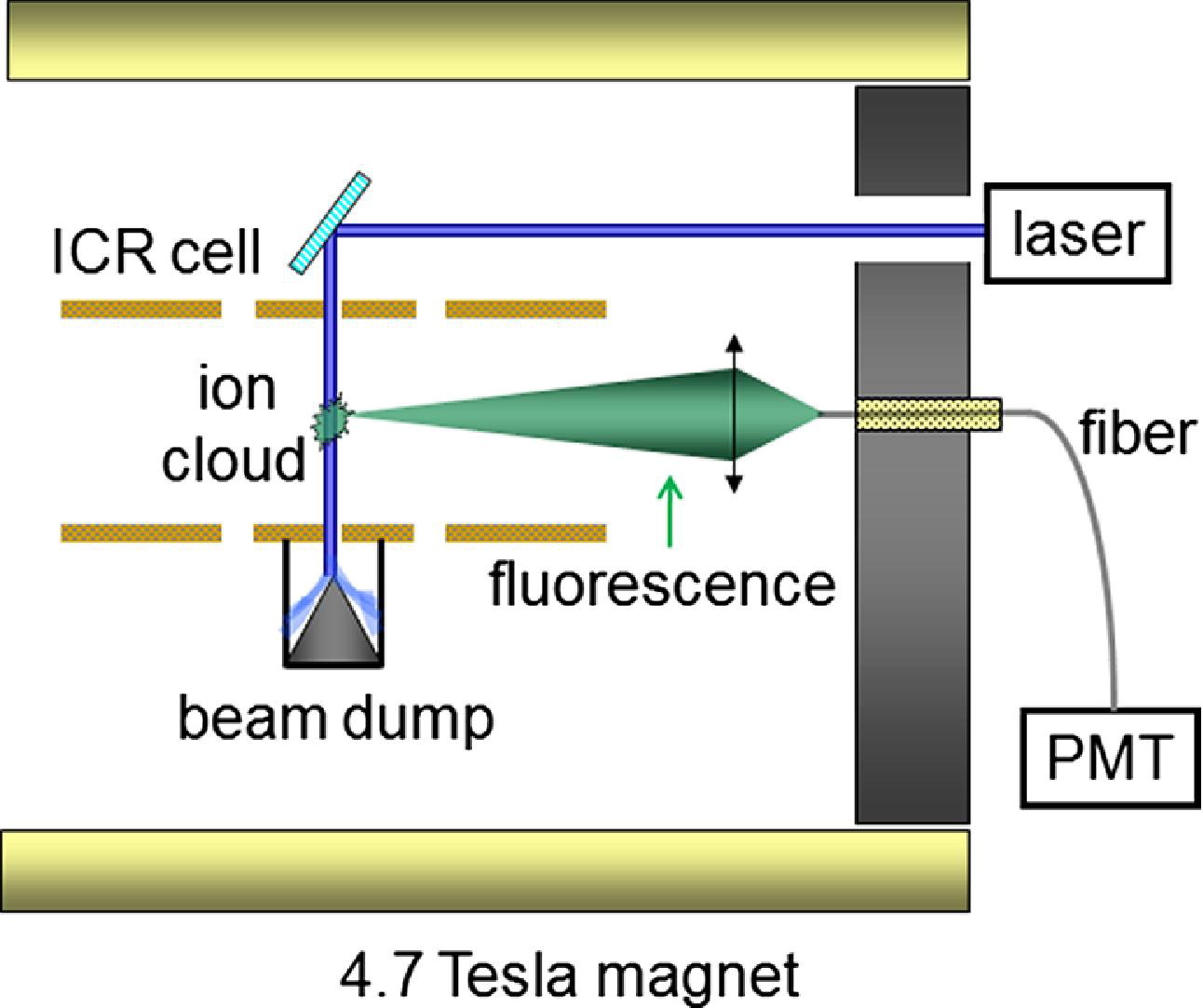}
\caption{Simplified schematic of the Fourier transform ion cyclotron resonance (FT-ICR) mass spectrometer setup employed by Chingin et al. \cite{chingin2009exploring}. It is based on the laser excitation of the ion cloud trapped in a Penning trap placed between two superconducting magnets and the fluorescence detection by an optics assembly and optical-fibre-coupled PMT. The system provided an overall detection efficiency of $1-2$\%. \textit{Adapted from \cite{chingin2009exploring} J Am Soc Mass Spectrom \copyright 2009 American Society for Mass Spectrometry. Published by Elsevier Inc.}.}
\label{figfticr}
\vspace{5mm}
\end{center}
\end{figure}

Renato Zenobi and co-workers \cite{chingin2009exploring} demonstrated a novel approach for the spectroscopic studies of trapped ions in the gas phase with an overall fluorescence detection efficiency of $1-2$\%. The schematic of this experiment is shown in figure \ref{figfticr}. They employed a Fourier transform ion cyclotron resonance (FT-ICR) mass spectrometer to explore the gas-phase fluorescence and photofragmentation of rhodamine 6G cations trapped inside a Penning Trap \cite{frankevich2005laser, chingin2009exploring, dashtiev2006effect}. The ions confined inside the trap, which sits between two superconducting magnets, were irradiated with an argon laser emitting at 488 nm, and the resulting fluorescence was collected orthogonal to the incident beam direction through an assembly consisting of collimating lenses, baffles, filters, and a fibre-coupled PMT for photon counting. The main challenge was photon detection in the limited space inside the trap. Still, the detection efficiency achieved using this setup ($1-2$\%)
was far better than those previously reported \cite{vandevender2010efficient, bian2010gas, sagoo2011fluorescence, friedrich2004time, forbes2011gas, sassin2009fluorescence}. Also, they probed the ions generated by a MALDI source as well as an electrospray ionisation source (ESI) \cite{frankevich2005laser, chingin2009exploring, yamashita1984electrospray}.

The gas-phase fluorescence resonance energy transfer (FRET) is an important application of fluorescence spectroscopy. Pioneered by Park and co-workers, this technique was further developed and implemented by Zenobi and co-workers for the investigating gas-phase FRET of bio-molecular ions \cite{danell2003fret}. Park used a quadrupole ion-trap-based setup to carry out the gas-phase FRET measurements of double-strand oligonucleotide anions, and Zenobi’s group used the FT-ICR spectrometer for the gas-phase FRET studies of rhodamine 6G-sulphorhodamine B FRET pair and carboxyrhodamine 6G-biodipy TR-FRET pair \cite{danell2003fret, dashtiev2005clear, frankevich2014fluorescence}. Jockusch and co-workers also reported gas-phase FRET measurements of polyproline peptides tagged with rhodamine dyes within a quadrupole ion trap \cite{talbot2010fluorescence}. The main challenges involved in these works were the low density of trapped ions, limited optical access, and careful choice of the FRET pair.

\begin{figure}[ht]
\begin{center}
\includegraphics[width=.6\linewidth]{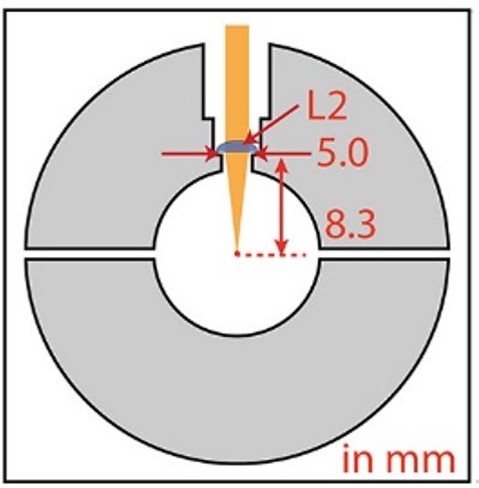}
\caption{Cross-sectional view of a customised quadrupole ion trap. Three holes are drilled on the ring electrode, two holes of diameters of 1.5 mm for providing laser access and the third hole of 5mm for collecting the emitted fluorescence. The collection lens, L2, is placed 8.3 mm from the trapped cloud to collect as much light as possible. The light collection efficiency of the system was 2.3\%. \textit{Adapted from \cite{tiwari2020breaking} Journal of American Society for Mass Spectrometry \copyright 2021 American Society for Mass Spectrometry. Published by American Chemical Society}.}
\label{figzen}
\vspace{5mm}
\end{center}
\end{figure}

In a very recent publication, Zenobi and coworkers demonstrated a technique called tm-FRET (transition metal fluorescence resonance energy transfer) in the gas phase with a relatively high fluorescence collection efficiency (2.3\%) \cite{tiwari2021transition}. Ions generated by a nano-ESI source were trapped in a customised QIT, probed with frequency-doubled Ti Sapph: laser pulses. The resulting fluorescence was measured by a spectrograph coupled with a CCD/single-photon avalanche photodiode. They modified a commercial LCQ mass spectrometer (Thermo electron corporation, USA) to improve optical access and fluorescence collection efficiency. For better access to the top lid of the LCQ, the printed circuit board, originally placed horizontally on the top lid to control the RF and DC voltages applied to the electrodes and the ion optics, was moved and fixed vertically. The modified LCQ trap is shown in figure \ref{figzen} \cite{tiwari2020breaking}. Two holes of diameter $1.5\,$mm were drilled on the ring electrode for providing laser access to the trapped ion cloud. A third hole of $5\,$mm was drilled for fixing a large numerical aperture (NA) lens for collecting the fluorescence light emitted orthogonal to the incident beam direction.

The trapped ion cloud had an ellipsoid geometry of length of about $1\,$mm and breadth less than $0.5\,$mm. The tm-FRET technique involves measuring the fluorescence lifetimes of rhodamine 110 dye in the presence and absence of a transition metal ion (Cu$^{2+}$). Rhodamine 110 acts as the electron donor while Cu$^{2+}$ acts as the electron acceptor. Donor fluorescence quantum yield decreases in the presence of an acceptor. The energy transfer efficiency between the donor and acceptor, known as the FRET efficiency, can be calculated from the measured fluorescence lifetimes. This technique allows one to perform intramolecular distance measurements in terms of fluorescence lifetime measurements. 

The original designs of the Paul trap and the Penning trap do not provide easy optical access. In this context, Vaishnavi Rajagopal et al. \cite{rajagopal2017linear} have explored the possibility of using a linear quadrupole ion trap that provides better optical access. They introduced an expanded inner diameter linear ion trap (eidLIT) with an inscribed diameter larger than that is used for an ideal linear quadrupole ion trap. This design allows them to perform the gas-phase fluorescence spectroscopy of low quantum yield fluorophores. The geometry of their trap was similar to the one shown in figure \ref{figql} but with a larger inscribed diameter ($\rho_0 = 0.8 r_0$ instead of $1.148r_0$) \cite{gerlich1992inhomogeneous,march2009quadrupole}. A continuous-wave solid-state laser emitting at $488\,$nm with a beam diameter of $2\,$mm irradiated the trapped ion cloud for $5\,$s. The fluorescence emission was collected through a $25.4\,$mm diameter lens of NA $0.6$, mounted $13.5\,$mm away from the eidLIT axis. The light emitted via fluorescence was focused on a fibre-coupled PMT from Hamamatsu. This apparatus enabled spectral acquisitions with an SNR of $100$ for the gas-phase rhodamine 6G and could measure the fluorescence lifetime of chromophores with quantum yields less than 10\%.

\begin{figure}[ht]
\begin{center}
\includegraphics[width=0.6\linewidth]{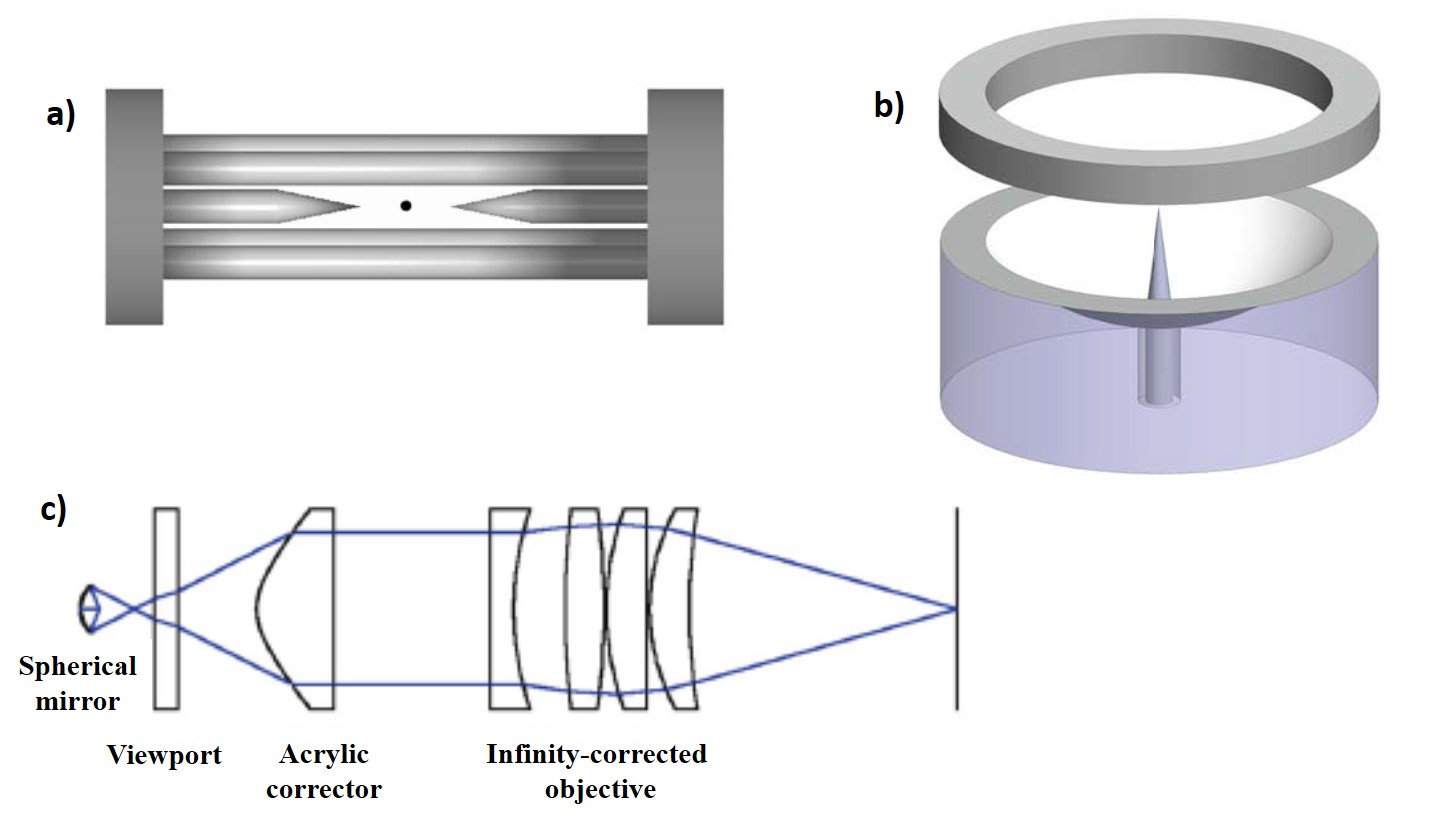}
\caption{A schematic of the novel tack trap. a) The first design of the linear Paul trap comprising of 4 quadrupole rods of 0.5 mm diameter and 1 mm spacing. The two tapered endcap needles are separated by 2.6 mm. b) The final tack trap design incorporated a large NA metallic spherical mirror with high reflection coating as one of the RF trap electrodes itself to maximise the solid angle of fluorescence signal collection and a sharp tapered needle as another electrode. Ion trapping occurs near the tip of the needle. An additional ring electrode was kept to increase the trapping depth. c) An aspheric corrector and a microscope objective for imaging the fluorescence after correcting the spherical aberrations caused by the large NA mirror. The trapped ions are irradiated by an extended cavity diode laser (ECDL) source, and fluorescence collection is performed orthogonal to this incident beam direction with a collection efficiency of 24\%. \textit{Reprinted from \cite{shu2011efficient} Journal of Optical Society of America B \copyright 2011 Optical Society of America}.}\label{figtt}
\vspace{5mm}
\end{center}
\end{figure}

Several groups adopted different geometries for maximising the solid angle of collection and hence the fluorescence collection efficiency from trapped ions \cite{vandevender2010efficient, shu2011efficient, shu2010efficient}. Vandevender et al. \cite{vandevender2010efficient} proposed an apparatus for fluorescence measurements of $\mathrm{^{24}Mg^+}$ ions trapped in a surface electrode Paul trap integrated with an optical fibre-coupled PMT. This enabled a good collection efficiency of 2.1\%, mainly limited by the NA of the collecting fibre. By improving the NA of the collection optics, one can improve the collection efficiency. In 2011, Blibinov and co-workers \cite{shu2011efficient} set a record value of the collection efficiency by adopting a novel approach for fluorescence detection. They employed an Al-coated spherical mirror geometry of high NA for one of the trap electrodes to reach 24\% of solid angle of collection, which is at least an order of magnitude higher than that could be achieved in previous works \cite{sagoo2011fluorescence, friedrich2004time, forbes2011gas}. This so-called tack trap was used to probe $\mathrm{{^{138}}{Ba}^+}$ ions \cite{blinov2006broadband, shu2011efficient, shu2010efficient}. A schematic of the fluorescence collection with the tack ion trap is shown in figure \ref{figtt}.
The first version of the trap is shown in figure \ref{figtt}(a), which is a linear Paul trap with 4 rods of $0.5\,$mm diameter separated by 1 mm, while the separation of the two endcap needles is about 2.6 mm. The fluorescence emission was collected by a lens and was directed to an image intensified camera for photon counting. In the second phase, this trap was improvised by mounting a large NA spherical mirror below the linear Paul trap with the mirror's focal point about 50 mm away from the trapping region. Laser-cooled $\mathrm{{Ba^+}}$ ions were trapped, and imaged with an electron-multiplied CCD camera \cite{shu2009trapped}. This geometry provided a collection efficiency of 11\%. However, a fraction of the fluorescent photons generated from the ions after being probed by the laser were blocked by the quadrupole rods of the trap. This limitation was later overcome by a novel trap design. In the so-called 'tack trap' (figure \ref{figtt}(b)), a large NA metallic spherical mirror (NA$\approx0.82$) coated with high reflective material itself acted as one of the trap electrodes. A 0.5 mm needle penetrating through the 0.75 mm hole drilled on the spherical mirror surface acted as the second trap electrode, and a ring electrode placed above the mirror acted as the third. Trapping and imaging of $\mathrm{{Ba^+}}$ ions were performed with a collection efficiency of 24\%, which is the highest value obtained in any of the fluorescence imaging techniques so far, the efficiency being limited by the mirror imperfections. In addition to this, an aspheric corrector was used to correct the spherical aberrations caused by the large NA metallic mirror. While this specialized design has been tested only with atomic ions, implementing it for molecular ions is not straightforward. However, advancements in ion-trapping strategies might enable the use of such traps for molecular ions in future.

In addition to the ambient temperature studies on fluorescence, a few research groups have addressed the low-temperature dependence of fluorescence characteristics of the gas-phase ions \cite{kordel2010laser,vogt2021effect, kjaer2021new,greisch2013intrinsic,danell2003fraying}. The molecular species of interest are cooled to low temperatures through a method referred to as the \emph{buffer-gas cooling} \cite{gerlich1992inhomogeneous}. In this process, an ideal gas such as helium is filled into the ion trap at relatively high pressure (of the order of a millibar), where the background pressure is typically below $1\times10^{-7}\,$mbar. The ion trap is maintained at a low temperature with the help of a cryostat. The buffer gas attains the temperature of the trap by collisions with the walls of the trap, and the ions get thermalised in all degrees of freedom with the buffer gas after a few collisions. Here we provide a brief overview of a few representative publications dealing with low-temperature fluorescence spectroscopy of xanthene-based dye molecules. In 2010, Schooss et al. reported the cold ion fluorescence spectroscopy of rhodamine cations in a Paul trap at $90\,$K \cite{kordel2010laser}. They measured the integrated LIF of rhodamine 6G cations ($\rm R6G^+$) at an excitation wavelength of $488\,$nm and estimated a lower bound to the triplet excited state of $\rm R6G^+$ to be around $2\,$s. They were also able to achieve more than 99\% of the $\rm R6G^+$ in the triplet state by optical pumping. Such studies offer opportunities for state-selective probing of molecular ions. Recently Steen Brondsted Nielsen and co-workers carried out cold-ion fluorescence measurements with a cylindrical Paul trap-based apparatus named LUNA-2 (Luminescence Instrument in Aarhus-2) \cite{vogt2021effect, kjaer2021new}. The fluorescence emission spectra of pyronin cation (PY$^+$), resorufin anion (R$^-$), rhodamine cations R575$^+$ and R640$^+$ were measured at a trap temperature around $100\,$K. They found that the fluorescence emission spectra of these molecular ions are blue-shifted and become narrow at low temperatures, in agreement with earlier work \cite{greisch2013intrinsic}.

The detection schemes adopted in the works presented so far fall in the single-pass detection schemes, i.e., the exciting photons interact with the trapped ion cloud in a single pass, and the collected fluorescence is collimated, filtered, and sent to a fibre-coupled detector. The path lengths, namely the ion-photon interaction path length and the ion-detector path length, were restricted to centimetre ranges which in turn limit the SNR while probing the weak fluorescence signals from gas-phase ions.
\subsection{Cavity-enhanced fluorescence spectroscopy}
Several groups explored the possibility of enhancing the SNR of the measured fluorescence signal to detect weak fluorescence emission. One way to achieve this is to increase the laser power. But increasing the power not only introduces additional amplitude noise associated with the source but also triggers other processes like multiphoton absorption, photoionisation, and photodissociation. An alternative could be incorporating a high finesse optical cavity in the LIFS instrumentation. The cavity could be placed either in the path of the photons exciting the fluorescing molecules or along the fluorescence detection axis. The former serves to enhance the photon-ion interaction path length by several orders of magnitude depending on the reflectivity of the cavity mirrors, which in turn provides the possibility of increased fluorescence emission, thereby enhancing the SNR. The latter involves enhancing the emitted fluorescence by coupling it to a high finesse optical cavity. The SNR enhancement levels are proportional to $2F/ \pi$, determined by the cavity finesse $(F)$, a parameter which is dependent on the reflectivity of the end mirrors of the cavity \cite{gagliardi2014cavity}. The higher the reflectivity, the higher the finesse, and therefore, the higher will be the SNR. A few studies were carried out by incorporating high finesse optical cavities in the LIFS measurements \cite{ross2007cavity, bixler2016utilizing, sanders2018absolute}. But all of them focused on fluorescence measurements in solution or neutral gas-phase molecules stored in a sample cell. To the best of our knowledge, only one work demonstrated successful integration of optical cavities in ion-trap setups \cite{benito2015optical}. One of the challenges in cavity integration is the space constraint due to the geometry of the conventional ion traps. However, it is indeed possible to integrate optical cavities with unconventional ion trap geometries such as eidLIT \cite{rajagopal2017linear} and wire traps \cite{rajeevan2021numerical,geistlinger2021sub}.

Ross et al. \cite{ross2007cavity} employed a method for measuring the molecular iodine fluorescence by placing the sample cell in an X-shaped, four-mirror cavity ($\approx$ 1.2 m pathlength) inside which the laser-sample interaction took place. This geometry enabled them to gain an order of magnitude in the SNR of the detected fluorescence. Bixler et al. \cite{bixler2014ultrasensitive, bixler2016utilizing} introduced a clever technique to enhance both the excitation and emission pathlength by placing the sample inside a spherical cavity in which the light-sample interaction took place. The cavity's inner surface was coated with a diffusively reflecting material. With this approach, in addition to having increased excitation pathlength, isotropic illumination allowed the fluorescent signal to be generated in the entire volume as well as the possibility to collect it from all $4\pi$ steradians, which in turn led to an enhancement in the collected fluorescence SNR \cite{ bixler2016utilizing}. The technique was successfully implemented for ultra-sensitive detection of trace amounts of urobilin, a waste product in water (down to femtomolar levels), and pyranine in the gas phase.

\begin{figure}[ht]
\begin{center}
\includegraphics[width=.6\linewidth]{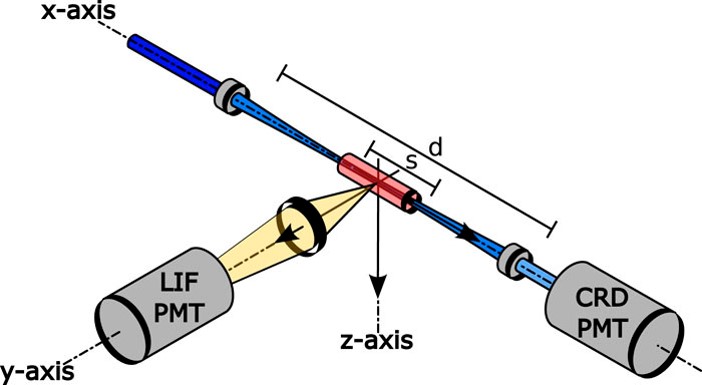}
\caption{A schematic of the cavity-enhanced laser-induced fluorescence (CELIF) apparatus. This technique is used to perform simultaneous measurements of LIF transient and CRD times from the same sample corresponding to a single pulse of laser excitation. Sample volume denoted by s, located in the ringdown cavity formed by two high-reflectivity mirrors separated by a distance d, is excited by the laser source along the cavity axis (x-axis), and the fluorescence emission is collected along the y axis. The CELIF signal, the ratio of the measured LIF transient and the CRD transient, is used to retrieve the fluorescence lifetime and quantum yields. \textit{Reprinted from \cite{sanders2018absolute} \uppercase{THE JOURNAL OF CHEMICAL PHYSICS} \copyright 2018 \uppercase{aip} publishing}.}\label{figcelif}
\vspace{5mm}
\end{center}
\end{figure}

Sanders et al. \cite{sanders2018absolute} demonstrated the cavity-enhanced laser-induced fluorescence (CELIF) technique which is based on simultaneous measurement of cavity ring-down (CRD) time and the laser-induced fluorescence with a single laser pulse. The resulting signal is called the CELIF signal, which is the ratio of the measured LIF transient and the CRD transient. A schematic of their apparatus is shown in figure \ref{figcelif}.

The CELIF apparatus combines well-known cavity ring-down spectroscopy (CRDS) and laser-induced fluorescence spectroscopy (LIFS). In a typical CRD experiment, light enters the cavity formed by two high reflectivity mirrors ($R \approx 0.999$) and gets reflected back and forth repeatedly. A small fraction of light leaks through the output mirror during each pass, which is measured with a PMT. The time required for the light intensity to decay to $(1/e)^\mathrm{th}$ of its initial value in the absence and the presence of an absorbing sample are called empty cavity ring-down time $(\tau_0)$ and ring-down time $(\tau)$, respectively, which are used to retrieve the absorption coefficient of the sample. The high finesse cavity used for the CELIF measurements has $\tau_0 \approx$ 1.2 to 2 $\mu$s and $F \approx 3200$. The fluorescence resulting from the interaction of the laser pulse and the sample is measured by the LIF setup installed orthogonally to the incident beam direction. The CELIF technique involves recording the transients from both LIF and CRD simultaneously on a single pulse shot with identical PMTs and normalising the LIF transient with the CRD transient. This normalised signal ($S^{CELIF}$) is independent of the intensity fluctuations of the laser. Here the optical cavity acts as a filter to suppress the background noise by 100. The $S^{CELIF}$ signal measurement process is a background-free or an absolute LIF measurement technique. The measurement process consists of measuring the time-integrated fluorescence transient ($S^{LIF}$) and time-integrated ring-down transient ($S^{CRD}$), which are given by the following equations.
\begin{equation} \label{eqlif}
S^{LIF}(\lambda) = \alpha(\lambda)\times \Phi_f(\lambda)\times g\times I_L
\end{equation}
\begin{equation} \label{eqscrd}
S^{CRD} = \frac{I_{in}\times T}{2}
\end{equation}
where $\alpha$ is the absorption coefficient, $\Phi_f$ is the fluorescence quantum yield, $g$ is a geometry dependent factor of the detection system, $I_L$ is the integrated light intensity interacting with the sample within the LIF probe volume given by $I_L= (S^{CRD}\times (1+R))/T $, $I_{in}$ is the incident intensity on the cavity input mirror, R is the mirror reflectivity, and T is the mirror transmittance. For cavity mirrors of very high reflectivities (R $\approx$ 1), the fractional absorption $\ll$ 1 and $I_L$ takes the form $I_L = (S^{CRD}\times 2)/T$. $S^{CELIF}$ is then determined from the measured quantities using the relationship:
\begin{equation} \label{eqcelif}
S^{CELIF} = \frac{S^{LIF}}{S^{CRD}} = \alpha \times k
\end{equation}
where $k= (2\times \Phi_f\times g)/T$ is a proportionality factor. Equation \ref{eqscrd} is valid in an empty cavity where the fractional absorption per pass is equal to zero and $T=1-R \ll 1$. The instrument was tested by measuring the lifetime and the absolute fluorescence quantum yield of gaseous 1, 4-bis (phenylethynyl) benzene (BPEB), which was reported to have a fluorescence lifetime of $500\,$ps and a quantum yield of $0.58$ \cite{chu2004vibronic}. At Sandia National laboratories, Benito et al. realized a set-up in which a moderate finesse optical cavity (F $\approx$ 150) was integrated with a micro-fabricated surface linear ion-trap [MESA fab, Sandia national Labs] to enhance the light collection from the trapped ${\rm Yb^+}$ ion \cite{benito2015optical}. The single ion located at the cavity waist was interrogated with a laser emitting at 369 nm and the resulting photon was collected by dedicated optics and PMT.
\section{Application towards the intrinsic fluorescence studies of weakly fluorescing molecules: Fluorescein as a model system}

For a molecule to be used for biomedical applications, it must have a high fluorescence quantum yield and be non-reactive to the environments of interest while maintaining easy conjugation with biological molecules. An excellent choice for this purpose is fluorescein \cite{zhao2019application, mcqueen2010intrinsic}. It is a synthetic, organic dye molecule with a remarkably high fluorescence quantum yield and therefore used in a wide variety of applications such as to label proteins in biochemistry, as a gain medium in certain dye lasers, and in forensics \cite{mcqueen2010intrinsic, klonis1996spectral, yao2013fluorescence}. Of its seven prototropic forms, the doubly deprotonated form (dianion) is responsible for the molecule’s high fluorescence quantum yield (0.93). However, the high quantum yield of this molecule is only limited to an aqueous environment \cite{mcqueen2010intrinsic, yao2013fluorescence, forbes2011gas}, and no fluorescence emission could be detected in the gas phase. Recent studies concluded that the dianion loses fluorescence in the gas phase because the energy of the exciting photon is used to remove an electron from the molecule instead of fluorescence emission \cite{mcqueen2010intrinsic}. However, there are still several unanswered questions on this molecule's fluorescence characteristics, such as the electron dynamics within the molecule during photodetachment, how the solvent environment stabilises the electron against photodetachment, etc. No in-depth study on this species has been conducted in the gas phase and controlled ion-solvent interaction either experimentally or theoretically. In fact, there is another example of a molecule exhibiting similar behaviour as that of fluorescein dianion, namely, the monoanion of p-hydroxybenzylidene-2,3-dimethylimidazolinone (HBDI), which is frequently used by several research groups as a model system for the chromophore of the wild-type green fluorescent protein (GFP). The monoanion of HBDI does not fluoresce in the gas phase because of the fast internal conversion from the excited state \cite{svendsen2017origin}. However, Svendsen et al. showed that the molecule becomes fluorescent at a low temperature (100\,K) when it is trapped in the excited state long enough to make it fluorescent.

Recent developments in the laser-induced fluorescence spectrometry techniques have enabled the scientific community to perform detailed investigations on the fluorescence properties of fluorescein in solution for its solvatochromic effects \cite{mcqueen2010intrinsic, sjoback1995absorption, gallagher2013temperature}. This section reports some selected experiments performed towards understanding the fluorescence properties of solvent-phase and gas-phase fluorescein and how the current advances in the LIFS instrumentation can be exploited for further investigations in this direction. Most gas-phase measurements were carried out using QIT-based instrumentation, while the solvent dependence measurements were performed with a SPEX Fluorolog tau2 fluorometer.

\begin{figure}[ht]
\begin{center}
\includegraphics[width=0.6\linewidth]{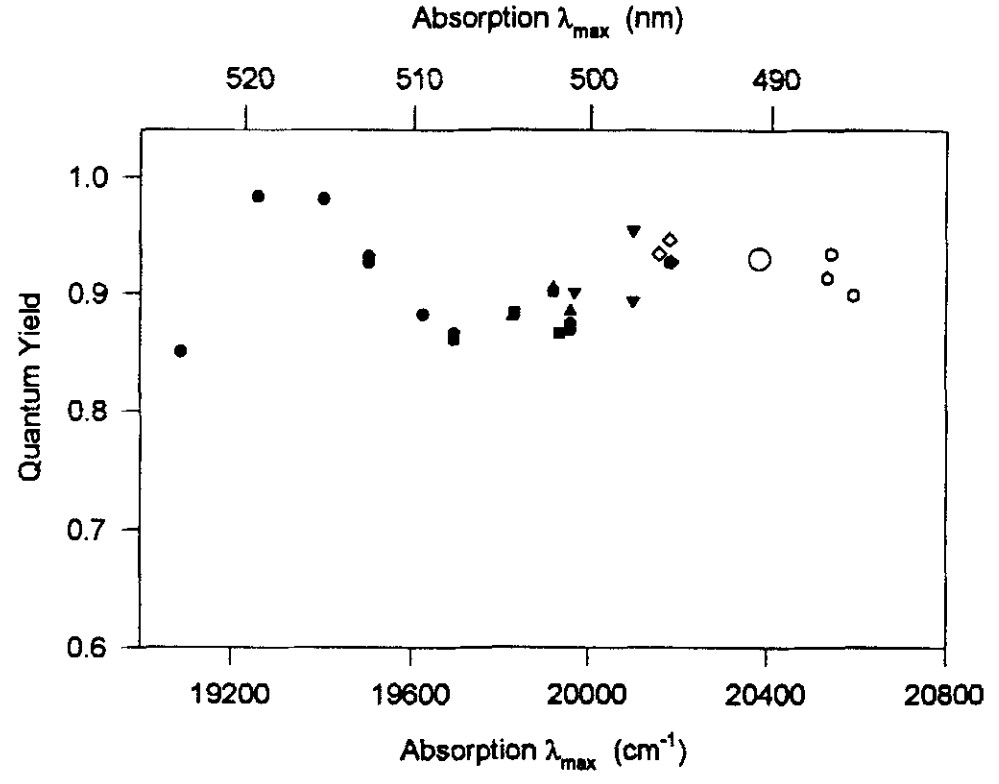}
\caption{Fluorescence quantum yield of dianion as a function of excitation wavelength with water and with various co-solvents. The quantum yield is only weakly dependent on the co-solvents ($\Phi_f \sim$ 0.93). \textit{Adapted from \cite{klonis1998spectral} Photochemistry and Photobiology \copyright 1998 American society for Photobiology}.}\label{figklonis}
\vspace{5mm}
\end{center}
\end{figure}

Fluorescein exhibits strong solvatochromism, a term coined to explain the fluorescence spectral dependence on solvent polarities. Klonis and co-workers carried out an experimental study to quantify the quantum yield dependence of fluorescein monoanion and dianion on solvent polarities \cite{klonis2000effect, klonis1998spectral, klonis1996spectral}. Absorption spectra were measured with a Cary 5 UV-VIS spectrophotometer, while the fluorescence spectra were measured with a SPEX Fluorolog Tau-2 fluorometer. Spectral acquisitions were performed by preparing fluorescein dianion and monoanion in water and with various co-solvents. The experimental outcomes showed that fluorescein monoanion quantum yield significantly depended on the solvent polarities, the measured value being minimum in water (0.36) and maximum in methanol (0.49).
One of the factors influencing the fluorescence quantum yield is the strength of the hydrogen bonding of the environment surrounding the fluorophore. Originally this observation was made by Martin et al., who demonstrated this effect unambiguously, and later on, several works continued exploring this effect \cite{ martin1975hydrogen, kamlet1979linear, taft1976solvatochromic, naderi2016solvatochromism, cramer1978hydrogen, rodgers1981picosecond}. For instance, the fluorescence quantum yield of fluorescein monoanion was reported to increase due to an increase in the hydrogen bonding of the co-solvent \cite{klonis1998spectral}. Several other works reported the fluorescence quenching of aromatic chromophores, namely, coumarin 102, fluorenone, and 2-naphthol, in the presence of water molecules caused due to the rupture of chromophore-solvent hydrogen bond in the excited state \cite{dobretsov2014mechanisms, turner1968quantitative}. Figure \ref{figklonis} depicts the fluorescence quantum yield dependence of fluorescein dianion in water and in various co-solvents. It is evident that the dianion fluorescence quantum yield does not depend significantly on the solvent environment. It is worthwhile to note at this point that the dianion loses its fluorescence in the gas phase.

In one of the recent studies, Gerasimova et al. investigated the fluorescence spectra for all protolytic forms of fluorescein (dianion, monoanion, neutral quinoid, neutral zwitterion, and cation) in a wide pH range (0.3 – 10.5) by combining experimental and computational methods \cite{gerasimova2020fluorescence}. Absorption and emission spectral acquisitions were carried out using a Lamba 35 spectrophotometer and Furolog 3-22 spectrometer, respectively, while the computations were based on time-dependent density functional theory (TD-DFT) \cite{cossi2001time, tomasi2005quantum, schmidt1993general}. Fluorescence peaks of monoanion (496 nm), neutral quinoid (550 nm), and zwitterionic (483 nm) forms were assigned. The positions of these peaks were not clear prior to this work. The ground state and excited state dissociation microconstants were also determined by the Förster cycle method \cite{shah1985kinetic} to distinguish ionic equilibriums of various tautomeric forms of fluorescein. This study provided a time-resolved picture of the interplay between the excited state proton transfer and fluorescence spectra of protolytic forms of fluorescein. The different experimental parameters for various LIFS experiments on the prototropic forms of fluorescein and its derivatives are reported in table \ref{tab:qy-lt}. The reported quantum yields and lifetimes belong to the ions in the solution form \cite{yao2013fluorescence, spagnuolo2009photostability, seely1988estimation, alvarez2001fluorescein, klonis1996spectral, martin1975hydrogen, magde2002fluorescence, orte2005three}.
\begin{table}[h]
\small\sf\centering
\caption{Parameters obtained from various experiments performed on fluorescein (FL), 2, 7 Difluorofluorescein (DFF), and 2, 7 Dichrolofluorescein (DCF) using LIFS. The reported quantum yields and lifetimes belong to the respective ions in the solution form. Rhodamine derivative (Rh 640) was chosen as the reference for relative quantum yield measurements. Numbers in the bracket indicate the references from which these values were taken.\label{tab:qy-lt}}
\begin{tabular}{llll}
\toprule
Molecule&$\lambda_{ex}(nm)$&$\Phi_f$&Lifetime (ns)\\
\midrule
FL dianion& 490 \cite{yao2013fluorescence,martin1975hydrogen,klonis1996spectral}&0.9 \cite{martin1975hydrogen}&4.3 \cite{magde2002fluorescence,alvarez2001fluorescein}\\
FL monoanion&450 and 470 \cite{yao2013fluorescence, martin1975hydrogen,klonis1996spectral}&$-$&3.7 \cite{alvarez2001fluorescein}\\
DCF dianion&502 \cite{spagnuolo2009photostability}&0.94 \cite{spagnuolo2009photostability}&$-$\\
DCF monoanion&465 and 490 \cite{spagnuolo2009photostability}& $-$&5 \cite{yao2013fluorescence}\\
DFF dianion&490 \cite{orte2005three}& 0.97 \cite{seely1988estimation}&$-$\\
DFF monoanion&450 and 470 \cite{orte2005three}&$-$&5 \cite{yao2013fluorescence}\\
\bottomrule
\end{tabular}\\[8pt]
\end{table}

\begin{figure}[ht]
\begin{center}
\includegraphics[width=1\linewidth]{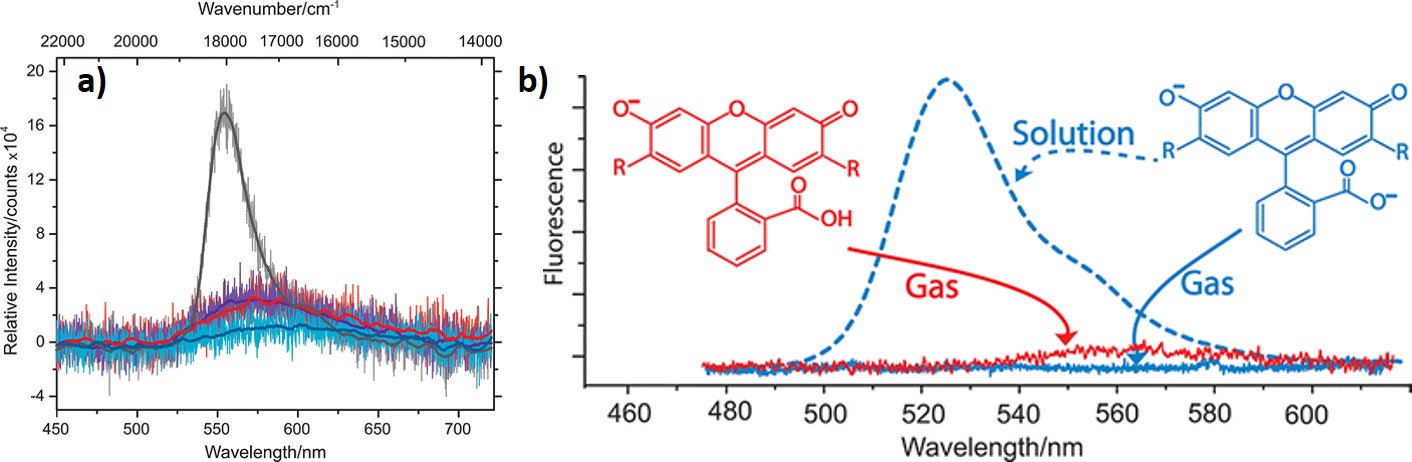}
\caption{a) Gas-phase fluorescence emission of monoanions of FL (blue), DFF (red), DCF (violet) and rhodamine 640 (grey). b) shows the fluorescence emission of FL dianion (solid blue trace) and monoanion (red) in the gas phase, while the blue dashed line shows the dianion fluorescence in solution-phase. \textit{Reprinted from \cite{yao2013fluorescence} THE JOURNAL OF PHYSICAL CHEMISTRY A \copyright2013 American Chemical Society}.}\label{figyao}
\vspace{5mm}
\end{center}
\end{figure}

Only a limited number of experimental studies have been reported on gas-phase fluorescein \cite{yao2013fluorescence, mcqueen2010intrinsic, kjaer2020gas, yao2011infrared, tanabe2012molecular, tiwari2020breaking}. Most measurements are either photodissociation spectroscopy of the molecule or photoelectron spectroscopy. To the best of the authors' knowledge, there are only three experiments on the gas-phase fluorescence measurement of any prototropic form of this molecule \cite{yao2013fluorescence,kjaer2020gas, tiwari2020breaking}. One of the possible reasons for the scarcity of such measurements is that the brightest prototropic form of fluorescein, namely, fluorescein dianion has been reported to have zero or very low fluorescence quantum yield in gas-phase \cite{mcqueen2010intrinsic}. Measuring very weak fluorescence signals can be challenging with the currently available single-pass fluorescence detection schemes due to SNR limitations. Yao et al. studied the spectral properties of fluorescein experimentally in the gas phase and the solvent phase \cite{yao2013fluorescence}. They measured the fluorescence emission spectra of monoanionic forms of fluorescein (FL), dichlorofluorescein (DCF), and difluorofluorescein (DFF) and compared them with that of rhodamine 640 \cite{yao2013fluorescence}. The experimental setup used was the same as that was shown in figure \ref{figbian} \cite{bian2010gas}. Trapped monoanions and dianions were irradiated for 800 ms with a laser operating at 510 nm, and the fluorescence emission was recorded orthogonally through a hole drilled on the ring electrode. The gas-phase fluorescence spectra of fluorescein and its derivatives measured by Yao et al. are shown in figure \ref{figyao}.

Rhodamine 640 showed the brightest fluorescence ($\Phi_f\approx 1$) in the gas phase \cite{yao2013fluorescence}. The fluorescence quantum yields of the monoanionic forms of fluorescein derivatives follow the order DCF $>$ DFF $>$ FL while FL dianion fluorescence was measured to be zero, or below the detection limit of the apparatus. The null-fluorescence of the gas-phase fluorescein dianion despite high quantum yields in solution was attributed to the electron photodetachment, which seems to outcompete fluorescence as a deactivation pathway. \cite{mcqueen2010intrinsic}. The null-fluorescence could also result from poor SNR of the setup, which does not have the sensitivity to detect the weak fluorescence signal.

Jockusch and coworkers established that the gas-phase fluorescence emission spectra of fluorescein monoanions are broad and red-shifted in contrast to the narrow blue-shifted spectra that they exhibit in the aqueous solutions \cite{mcqueen2010intrinsic, yao2013fluorescence}. One reason for this could be the internal proton transfer that occurs between the carboxylic acid groups and one of the xanthene oxygens in the excited state, as proposed by Kj\ae r et al. \cite{kjaer2017sibling}. However, Verlet and co-workers argued that this broad, red-shifted spectrum could be attributed to the partial charge transfer between the xanthene and the carboxyphenyl groups, occurring due to the rotation of the carboxyphenyl group relative to the xanthene group \cite{horke2015time}. To cast light on this problem, Kj\ae r et al. attempted to perform the gas-phase action and fluorescence spectroscopy of fluorescein monoanions and two derivatives where the carboxyl groups were replaced with an ester and a tertiary amide group \cite{kjaer2020gas}. Deprotonated forms of fluorescein monoanion, fluorescein methyl ester, and fluorescein piperidino amide were investigated. Fluorescence measurements were carried out by LUNA (luminescence instrument in Aarhus), whereas the photo-induced dissociation measurements were performed at an accelerator mass spectrometer \cite{stochkel2011absorption}, the details of which are explained elsewhere \cite{wyer2012absorption, stockett2016cylindrical, stockett2016nile}. Briefly, the ions trapped in a cylindrical Paul trap were excited with an Nd: YAG laser at 487 nm through a hole drilled on the cylindrical electrode. The exit electrode had a mesh grid geometry through which the fluorescence was collected with a spectrograph and CCD. This mesh grid geometry of the exit electrode enabled maximum fluorescence collection efficiency. The recorded spectra were in good agreement with the previously reported works by Yao et al. and McQueen et al. \cite{ yao2013fluorescence, mcqueen2010intrinsic}. Deprotonated fluorescein methyl ester and monoanion still exhibited broad, red-shifted fluorescence emission spectra with band maxima of 595 and 605 nm, respectively. This result disproved the argument that the internal proton transfer was responsible for these broad red-shifted spectra. On the other hand, the deprotonated fluorescein piperidino amide exhibited a spectrum narrower than the others, with the band maximum shifted slightly to the blue region (575 nm). The rotational structure calculations of the fluorescein derivatives mentioned above showed that the rotational dynamics for the amide was slower than the other two, thereby accounting for the narrower, blue-shifted fluorescence spectra. Their results showed good agreement with Horke et al. \cite{horke2015time}, who employed time-resolved photodetachment anisotropy (TR-PA) measurement scheme to probe the rotational dynamics of complex molecules. Renato Zenobi and coworkers used a modified LCQ mass spectrometer (discussed in the previous section) for measuring fluorescence emission from the monoanionic forms of fluorescein \cite{tiwari2020breaking} and found close agreement with the spectrum reported by Yao and Jockusch \cite{yao2013fluorescence}.

Apart from these publications, no other experimental work has been reported to explore further the gas-phase fluorescence measurements of fluorescein monoanion and dianion, to the best of our knowledge. Considering the enormous amount of applications of fluorescein \cite{minta1989fluorescent,kojima1998detection, tanaka2001rational, burdette2001fluorescent, gaylord2003dna}, there is a need for more experimental and theoretical investigations of the gas-phase fluorescence and photodetachment of fluorescein dianion to quantify their contribution to the de-activation pathways and better understand their electronic action spectrum. With the latest developments in a traditional LIFS setup in general, such as incorporating high finesse optical cavities for enhancing the fluorescence emission and adopting novel detection schemes to have a maximum solid angle of collection, one can enhance the SNR and push the detection limits further to explore even the molecules having weakest fluorescence. By exploiting the huge potential of such techniques, a deeper understanding of the photoinduced processes associated with fluorescein and similar fluorophores can be achieved, which in turn could enable designing molecules with predictable fluorescence characteristics for targeted applications.

\section{Conclusions}

This review provides an overview of laser-induced fluorescence spectroscopic techniques employed for the lifetime and quantum yield measurements involving gas-phase molecular ions. Both theoretical and experimental pictures are provided based on a series of state-of-the-art LIFS setups dedicated to quantum yield and lifetime measurements of fluorescent molecules. The main challenges involved in such experiments are discussed, and how to incorporate the recent advances in the LIFS instrumentation to overcome these challenges are pointed out. Adopting optical cavity-enhanced techniques to amplify the signal-to-noise ratio and implementing novel geometries to maximise the solid angle of fluorescence collection have proven to be vital in progressing towards this goal \cite{sanders2018absolute, shu2011efficient,vandevender2010efficient}. Finally, the state-of-the-art experiments performed towards the intrinsic fluorescence studies of Xanthene-based dyes, fluorescein, in particular, are described. Only a few works reported the experimental determination of the weak gas-phase fluorescence emission from the monoanionic forms of fluorescein, while the only measurement of the gas-phase fluorescence emission from fluorescein dianion detected no fluorescence \cite{yao2013fluorescence}. By implementing cavity-enhanced techniques combined with novel trap geometries ensuring the large solid angle for light collection and high SNR, these instruments can be very well extended in the near future towards the intrinsic fluorescence studies of least-investigated weakly fluorescing biomolecules of interest, fluorescein being just one example.

\end{document}